# On the existence of ethylenediaminetetraacetic acid (EDTA) doped zinc sulphate heptahydrate crystal

Bikshandarkoil R. Srinivasan
Department of Chemistry, Goa University, Goa 403206, INDIA
Email: srini@unigoa.ac.in  Telephone: 0091-(0)832-6519316; Fax: 0091-(0)832-2451184

**Abstract**

It is argued that the ethylenediaminetetraacetic acid (EDTA) doped zinc sulphate heptahydrate crystal reported by Raja et al Spectrochim. Acta 99A (2012) 23 is the well-known zinc sulphate heptahydrate.

**Keywords:** zinc sulphate heptahydrate; ethylenediaminetetraacetic acid; dubious crystal

Zinc sulphate heptahydrate (**1**)[1] represented by the formula $[Zn(H_2O)_6]SO_4 \cdot H_2O$ occurs in nature as the mineral *goslarite* and is a member of the epsomite group of minerals, which crystallizes in the orthorhombic Sohncke space group $P2_12_12_1$. Recently the Peterson group has reported a detailed study of the crystal structure and H-bonding of a synthetic *goslarite* [1]. In their study these researchers have shown that the *goslarite* structure is intolerant to doping (substitution) by even small ions like Fe(II) or Cu(II) beyond a concentration of 0.01 mol. % of dopant. Hence, the report of the authors of [2] claiming the inclusion (doping) of a large organic molecule namely ethylenediaminetetraacetic acid (EDTA) (**2**) (Fig. 1) in the *goslarite* structure appeared very unusual. A perusal of the details in [2] revealed several inconsistencies, indicating that the report is erroneous as shown below.

*EDTA doped zinc sulphate heptahydrate (**3**) and ninhydrin test for EDTA*

From an aqueous solution containing **1** and **2** in 1:01 mole ratio in dilute $H_2SO_4$, crystals of **3** are claimed to have been grown by slow evaporation solution growth technique [2]. Although the authors state '*XRD report confirms that crystal belongs to the Monoclinic*' as one of the highlights of their paper, no X-ray structural features of this 'so called' semi organic NLO material, and no details of structure refinement and the XRD instrumentation can be found in the entire paper[2]. The exact molecular formula for **3** is not given anywhere in the report, and the only information about the chemical composition of **3** is that it contains ~188 μg of **2** per ml. This very strange composition was supposedly determined based on a colorimetric estimation of **2** in the single crystals of **3** using ninhydrin as the colour forming reagent. It is well documented that α-amino acids containing an amino group (–NH$_2$) react with ninhydrin to give a characteristic coloured product known as Ruhemann's purple (Fig. 2) [3, 4]. Since **2** (EDTA) does not contain any –NH$_2$ group, the reported claim of determination of **2** by the ninhydrin method is absurd. This result can be termed as dubious.

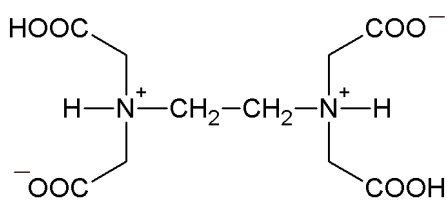

**Fig. 1** The zwitterionic structure of ethylenediaminetetraacetic acid (EDTA) **2**

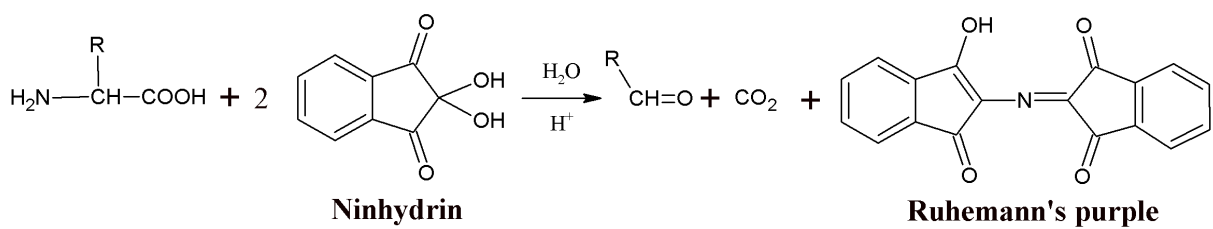

**Fig. 2.** Reaction of α-amino acid with ninhydrin. For mechanism of reaction see [4].

*Inconsistent synthetic and IR spectral details for EDTA doped zinc sulphate heptahydrate*

Authors of [2] claim to have grown the so called EDTA doped crystal **3,** by adding the dopant namely EDTA **2** to a saturated zinc sulphate heptahydrate solution and further adjusting the pH of the solution to 5.0 by addition of dilute sulfuric acid. The solubility of EDTA (mol. weight 292.24) in water is 0.5 g / L [5] and hence it is surprising to note that such a less soluble solid was chosen and used for crystal growth in water. It is well documented that the pH of saturated zinc sulphate heptahydrate solution is 4.5 [5]. In view of this, it is not clear how these authors could vary this pH to 5.0 (less acidic) by adding a strong acid like dilute sulfuric acid and grow the 'so called' doped crystals. The questionable nature of the synthesis of **3** can be evidenced from the experimental details of crystal growth where the authors state '*Under high acidity, growth rate decreases considerably and there may be a chance to form slightly soluble zinc hydroxides*'.[3] The only inference that can be made about the synthesis is that the reaction medium was strongly acidic with a pH value much less than 4.5.

In an attempt to prove the presence of EDTA in **3,** infrared data of compounds **1** - **3** were assigned albeit incorrectly in Table 1 of [2]. The authors did not consider the –OH stretching (~3540 cm$^{-1}$) and bending vibrations (~1620 cm$^{-1}$) of the water molecules of **1** as well as the EDTA doped compound **3**, while assigning two bands below 1000 cm$^{-1}$ for out of plane bending and librational modes for water in **1** and **3**. The intense vibration at ~ 3540 cm$^{-1}$ was assigned for N-H stretching vibration in **3** with an unacceptable explanation that the –NH stretching vibration is shifted to higher frequency due to complex formation indicating N-Zn bond. Any type of bonding to Zn by N can be ruled out from the structure of **2** (see Fig. 1) which carries a positive charge on the N atoms, making such N atoms unavailable for coordination to Zn(II). To justify the IR assignments, the authors cited the Nakamoto book[4] [6] and concluded that the four oxygens from carboxyl groups and two nitrogens from amino groups of EDTA would be coordinated to the central Zn ion as discussed[5] in [6]. This finding of the authors namely the hexacoordination of EDTA can also be ruled out as it is well known that **2** can function as a hexadentate ligand when it is a tetraanion and such an ion cannot form under the acidic reaction conditions employed for crystal growth. The binding of a small amount of **2** (dopant) to Zn(II) is contrary to all known principles of structural chemistry of solids, as this amounts to saying that in a single crystal**,** a major amount of Zn(II) exists as zinc sulphate heptahydrate with a very small amount of Zn(II) being coordinated to EDTA. The absence of a characteristic band at ~1698 cm$^{-1}$ [6] expected for the unionized –COOH group in the dopant **2,** in the reported IR spectrum gives clear evidence that no EDTA **2** is present in the grown crystal.

*The dubious dopant*

The characteristic 1698 cm$^{-1}$ signal is also not listed by the authors in the Table of IR data for the EDTA indicating that the authors did not use **2** as dopant. Although, the exact nature of the dopant used by the authors is not clear, the assignment of a band at 1619 cm$^{-1}$ in **3** for carboxylate ion stretching of the dopant is incorrect, as this signal can only be assigned for the bending vibration (–OH) of the water of the heptahydrate. In spite of all the above wrong assignments, the authors of [2] declared, '*Since our dopant is having same functional groups as that of glycine, we have compared* *with those references. Thus, the observed frequencies are in agreement with the available data and confirm that the grown crystals are EZSHH'*.

It is most inappropriate (and also unscientific) to confirm a crystalline compound based on a single IR spectrum as has been done in [2] in the absence of other supporting data. Such a conclusion is totally unacceptable for any compound especially for a crystal grown in a dubious manner. The only acceptable information from the reported IR spectrum (Fig 1 in [2]) is that the compound under study is zinc sulfate heptahydrate **1** as the spectrum has all the signals of **1** at the expected positions excepting a band at ~1460 due to some contamination, which can be attributed to some impurity based on the authors' own statement '*The influence of other important factors such as impurity, pH and supersaturation, on the crystal morphology may also be considered by analyzing their effects on the configuration of clusters and the bonding process within the boundary layer*' as quoted from the experimental section.

*Chemistry of growth of zinc sulfate heptahydrate crystals*

In their study, the Peterson group prepared a synthetic analogue of *goslarite* as 2 to 3 cm long acicular crystals by reaction of $ZnSO_4$ in $H_2O$ and 0.1 M $H_2SO_4$ (dilute sulfuric acid) and deuterated *goslarite* by using $D_2O$ and 0.1 M $D_2SO_4$ [1], which indicates clearly that zinc sulfate grows as long heptahydrate crystals (*goslarite*) in an acidic medium. Based on this chemistry, it can be correctly concluded that the authors of [2] who did a similar experiment using $ZnSO_4 \cdot 7H_2O$ in $H_2SO_4$ also ended up in crystals of **1** namely zinc sulfate heptahydrate but of of 4 mm length. The smaller size of the crystals of the authors of [2] can be attributed to the presence of impurities (dopant) in the crystal growth medium. Thus the formation of crystals of pure **1** without any dopant is well in accordance with the chemistry and reported structure of **1**.

In summary, it is shown that the recently reported "ethylenediaminetetraacetic acid (EDTA) doped zinc sulphate heptahydrate" is a dubious crystal.

**Footnotes**:

[1] In this paper, compounds are identified by numbers instead of non-standard abbreviations like ZSHH, EZSHH, LVZS.
[2] This is one of the several instances of the careless manner in which this work was described in [2].
[3] Zinc hydroxide cannot form in strongly acid medium.
[4] The book by Nakamoto was cited as Ref.21 in [2].
[5] There is no such assignment in the book by Nakamoto as reported by the authors of [2].